\documentclass[a4paper,fleqn]{cas-dc}

\usepackage[super]{natbib}
\def\tsc#1{\csdef{#1}{\textsc{\lowercase{#1}}\xspace}}
\tsc{WGM}
\tsc{QE}
\tsc{EP}
\tsc{PMS}
\tsc{BEC}
\tsc{DE}

\begin{document}
\let\WriteBookmarks\relax
\def\floatpagepagefraction{1}
\def\textpagefraction{.001}

\shorttitle{Tracking Dynamical Transitions using Link Density of Recurrence Networks}

\shortauthors{Rinku Jacob et~al.}

\title [mode = title]{Tracking Dynamical Transitions using Link Density of Recurrence Networks}                      

\author[1]{Rinku Jacob}
\cormark[1]
\ead{rinku.jacob.vallanat@gmail.com}
\affiliation[1]{organization={Rajagiri School of Engineering and Technology},
    addressline={Rajagiri Valley, Kakkanad}, 
    postcode={682039}, 
    country={India}}

\author[2]{R. Misra}[style=chinese]
% Email id of the second author
\ead{rmisra@iucaa.in}
\affiliation[2]{organization={Inter-University Centre for Astronomy and Astrophysics},
    addressline={Pune}, 
    postcode={411007}, 
    country={India}}
 
\author%
[3,4]{K P Harikrishnan}
\ead{kp.hk05@gmail.com}
\affiliation[3]{organization={Ramalayam, Chitoor Road},
     city={Cochin},
     postcode={682011}, 
    country={India}}

\affiliation[4]{organization={Previously at: The Department of Physics, The Cochin College},
    city={Cochin},
    postcode={682002}, 
    country={India}}    

\author%
[5]{G Ambika}
\ead{g.ambika@iisertvm.ac.in}
\affiliation[5]{organization={Indian Institute of Science Education and Research},
    city={Thiruvananthapuram},
    postcode={695551}, 
    country={India}}

\cortext[cor1]{Corresponding author}

\begin{abstract}
We present Link Density ($LD$) computed from the Recurrence Network (RN) of a time series data as an effective measure that can detect dynamical transitions in a system. We illustrate its use using time series from the standard R\"ossler system in the period doubling transitions and the transition to chaos. Moreover, we find that the standard deviation of $LD$ can be more effective in highlighting the transition points. We also consider the variations in data when the parameter of the system is varying due to internal or intrinsic perturbations but at a time scale much slower than that of the system dynamics. In this case also,  the measure $LD$ and its standard deviation correctly detect transition points in the underlying dynamics of the system.  The computation of $LD$ requires minimal computing resources and time, and works well with short datasets. Hence, we propose this measure as a tool to track transitions in dynamics from data, facilitating quicker and more effective analysis of large number of data sets.
\end{abstract}

\begin{highlights}
\item Link Density ($LD$) of Recurrence Networks computed from data is shown as an effective measure to track dynamical transitions 
\item Its efficiency in detecting transitions is illustrated in the context of bifurcation transitions and dynamical transitions under slowly varying parameter for the R\"ossler system. 
\item It requires minimal computational resources, and it is effective even with short datasets and hence offers an efficient method for analysing large datasets
\item The standard deviation of $LD$ is shown to indicate transitions between periodic and chaotic states.
\end{highlights}

% Keywords

\maketitle

\section{Introduction}
Understanding and predicting the dynamical behavior of complex systems is an important area of research across several scientific and engineering domains. Central to this endeavor is the concept of the variation of the control parameters of the system, which critically influences a system's state and evolution. The changes in these parameters can happen naturally, due to external influences or due to random fluctuations. However, such changes can cause transitions in the dynamics of the system, resulting in sudden or slow but significant qualitative changes in dynamics, like the transition from regular dynamics to chaos\cite{strogatz2018nonlinear,ott2002chaos}.  It is important but challenging to track under which conditions the state of a system loses stability and transit to different dynamics.  Most often dynamical transitions in complex systems are to be understood by analyzing observational or measured data or time series.  Then identifying a proper measure that can effectively detect changes in dynamics is essential for proper analysis and prediction.  The accurate detection of such changes can influence theoretical and practical outcomes in fields ranging from climate science\cite{donges2011identification,collins2013long} to bioinformatics\cite{karain2017detecting}. Moreover, anticipating transitions or other significant changes in system dynamics is often very crucial in diverse fields from economics\cite{shiller2015irrational} to epidemiology \cite{anderson1991infectious}.  In all these cases, the challenge lies in accurately and swiftly measuring the variability of data, particularly when data sets are limited.  It is crucial for making timely and effective predictions and for understanding of the system's dynamics and its response to varying control parameters. 

The methods of nonlinear time series analysis provide many tools for studying the characteristic features of complex dynamical systems from their time series.  In this context, the recurrence measures based on the quantification of the recurrence pattern of trajectory points in the phase space reconstructed from data are widely used.  Among them, recurrence plot measures computed using Recurrence Quantification Analysis (RQA), provide a package to analyse the dynamical state and its transitions \cite{marwan2007recurrence,marwan2013recurrence}. Thus the measure Lacunarity, is shown to identify changes in the dynamics\cite{braun2021detection} and studies in thermoacoustic systems underscore the effectiveness of RQA as early warning signals to predict critical shifts from chaotic to limit cycle states\cite{pavithran2021critical}. Further, the introduction of bootstrap techniques to the RQA process has enhanced the reliability of these methods by providing statistical confidence, essential for validating the detected changes  \cite{karain2017detecting,marwan2008significance, souza2008using}. 

The method of RQA is closely followed by measures from recurrence networks. Recurrence networks (RNs) transform time series data from dynamical systems into complex network structures by utilizing the concept of recurrence, a fundamental characteristic of bounded systems\cite{eckmann1995recurrence}. In actual practice, the recurrence matrix from the time series is used to get the adjacency matrix of the recurrence network, which links different points that are close neighbours in the reconstructed phase space.  Thus the recurrence networks derived from the time series fully characterize the geometric properties of the phase space of the underlying dynamical system \cite{donner2010recurrence}.  Since any transition in dynamical states results in changes in the recurrence structure, it can be detected from consequent changes in the recurrence network measures.

The advantages of recurrence network measures in identifying dynamical structures are clearly demonstrated in a few cases like transitions in logistic map \cite{marwan2009complex} with changing control parameter and a real-world paleo climatic time series\cite{donges2011identification}. Recently this has been applied in detecting dynamical transitions during human speech production \cite{lal2022recurrence}, stock market dynamics \cite{krishnadas2022recurrence}, and in a turbulent combustor\cite{kawano2023complex}.  Additionally, RN measures are known to work well with shorter datasets \cite{jacob2016uniform}.  Some of the recent trends in this direction are higher dimensional recurrence networks\cite{yang2019new} and directed recurrence networks\cite{delage2023directed, zhao2023robustness}.

In this study, we present the potential of recurrence network measures in the detection of dynamical transitions from data or time series. We consider the data from the R\"ossler system, with its well-documented transitions, and compute the related recurrence network measures to identify which ones are most sensitive to dynamical transitions and can be easily computed with short datasets.  This study extends the previous research by using network measures to investigate how changes in control parameter affect the system's dynamics.  Our study in this direction indicates that among the recurrence network measures, Link Density ($LD$) that can capture the properties of the phase space density of the underlying dynamics can be most effective as it quantifies the heterogeneity and variations in the distribution of points in phase space. Additionally, we try to see if this network measure can effectively track the dynamics of the system when parameters change slowly over an extended time.  This search serves as a preliminary step towards understanding the dynamical transitions in real systems where parameters can change due to various types of perturbations or external influences. 

\section{\label{sec:level1}Network measures from data}
The nature of dynamics underlying data and its variations are vital in the study of complex systems.  To estimate the relative efficiency of different approaches in this context, we take R\"ossler system as the model in our study.  It is a relatively simple model that can exhibit a broad spectrum of behaviors, from regular, predictable patterns to chaotic, unpredictable dynamics, based on the values of its parameters.  Its dynamics follows the set of equations: 

\begin{align}
\begin{split}
\frac{dx}{dt} &= -y - z, \\
\frac{dy}{dt} &= x + ay, \\
\frac{dz}{dt} &= b + z(x - c).
\end{split}
\label{eq:rossler}
\end{align}

We keep $a=b=0.2$ and vary the parameter $c$, to generate data corresponding to various possible dynamical states and explore the system's dynamics by analysing the data. The construction of a recurrence network (RN) from data of size $(N_T)$ starts with the time delay embedding of the scalar time series  $s(1),s(2),...,s(N_T)$ into an $M$ - dimensional space \cite{ambika2020methods,tan2023selecting}. This embedding process, essential for reconstructing the system's trajectory in its phase space, needs selecting an appropriate time delay $\tau$, which is taken as $1/e$ of the maximum value of autocorrelation function \cite{jacob2016uniform} in this work. As part of pre-processing, the time series is initially transformed into a uniform deviate before the time delay embedding. This transformation rescales the size of the embedded trajectory into a $M$-dimensional  unit cube $[0,1]^M$ \cite{jacob2016uniform}. In this study, we deal with data from R\"ossler system and hence $M$ is taken as its actual dimension 3. 

The embedding results in $N = N_T - (M - 1)\tau$ state vectors that collectively represent the trajectory. The proximity and local density of points on the trajectory are mapped into the structure of the recurrence network. For this, each point on the trajectory is treated as a node and a pair of nodes $i$ and $j$ are connected if their Euclidean distance in the reconstructed phase space is less than or equal to a specified recurrence threshold $\epsilon$.  This procedure generates the recurrence matrix $R$, with $R_{i,j}= 1$ indicating a connection between nodes $i$ and $j$, and $R_{i,j}= 0$ otherwise. The adjacency matrix $A$, which represents the RN, is derived from $R$ by excluding self-loops, thus,  $A \equiv R - I$, where $I$ is the identity matrix.  By this construction, $A$ is a binary symmetric matrix, reflecting the RN's nature as an unweighted and undirected network. The appropriate value of $\epsilon$ is chosen as that at which we get a single network \cite{jacob2016uniform}.  This is ensured by monitoring the second smallest eigenvalue $\lambda_2$  of the Laplacian matrix $L = D - A$, where $D$ is the diagonal degree matrix.  When $\lambda_2$  moves from negative to zero, it signifies that all nodes are part of a single connected network\cite{eroglu2014finding, jacob2016uniform}. A common range of threshold $\epsilon$ applicable to all-time series for a given embedding dimension $M$ is established in an earlier work \cite{jacob2016uniform}. When analysing many data sets, $\epsilon$ is then chosen as the minimum value within this range. For an embedding dimension of $M=3$, this minimum threshold is found to be $0.1$, which is adopted as the value of $\epsilon$ for the RN construction in our analysis.

In our study, we focus on three significant network measures\cite{donges2012analytical}: Link Density ($LD$), average Clustering Coefficient ($CC$), and Characteristic Path Length ($CPL$). Our study indicates that $CC$ and $CPL$ show less pronounced variations near dynamical transitions compared to $LD$(and hence not included here). The preference for $LD$ is also driven by its practical advantages, particularly its suitability for real-time applications and in scenarios with limited data availability. 
Calculating $LD$, which quantifies actual links relative to potential links within the network, is notably faster than computing $CC$ and $CPL$, making it an efficient choice for monitoring dynamics in response to variations in control parameter.  The $LD$ is calculated using the equation \cite{donges2012analytical},
\begin{equation}
LD = \frac{\sum_{i=1}^{N}\sum_{j=1}^{N} A_{ij}}{N(N-1)},
\label{eq:linkdensity}
\end{equation}

where \(A\) is the adjacency matrix,  \(N\) is the total number of nodes in the network. 
\section{\label{sec:level1}Bifurcation Transitions}
\begin{figure}[ht]
\centering
\includegraphics[width=0.5\textwidth]{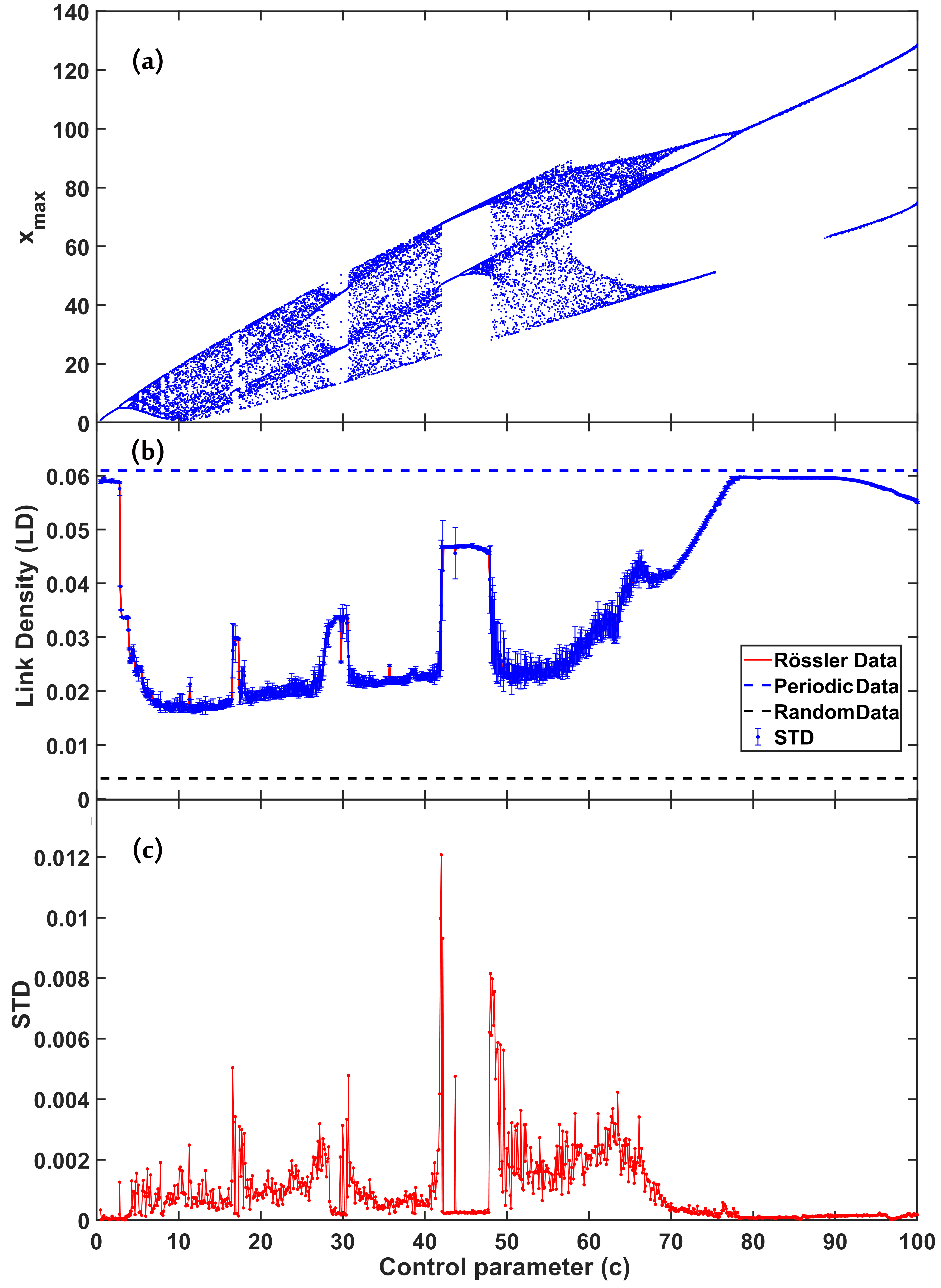}
\caption{Characterization of the transitions in the dynamics of the R\"ossler system derived from the time series (a) Bifurcation diagram showing the transitions between different dynamical states as parameter $c$ is varied within a range from 0.5 to 100, keeping $a=b=0.2$. (b) Variations in average $LD$ for the same range of $c$ values. The error bars indicate the standard deviation of $LD$ for each $c$ value. For comparison, the value of $LD$ from purely periodic data is shown by a blue dotted line and that of pure Gaussian white noise by a black dotted line. (c) Variations of the standard deviations of $LD$ values show an increase within the chaotic region hence effectively indicating points of transitions in dynamical states. }
\label{fig:fig1}
\end{figure}

We first apply the network measure $LD$, to study the bifurcation transitions in the R\"ossler system.  The time series data required for this is generated from the R\"ossler equation (\ref{eq:rossler}) with parameters $a$ and $b$ set to 0.2, while varying the third control parameter $c$ from 0.5 to 100 in increments of 0.1.  The bifurcations and transitions in the dynamics for this range of $c$ values can be identified from the bifurcation diagram plotted by direct calculations using the system's equations in Fig.~\ref{fig:fig1}a. For each value of $c$, the time series of length 50,000 is generated with a chosen set of initial values ($x_0=0.2, y_0=0.2, z_0=0.2$). After removing 5000 transient points, the data is divided into 15 sets, each comprising $N= 3000$ points. Each set is embedded in a 3-dimensional space and the recurrence pattern in the reconstructed trajectory is captured in its adjacency matrix. The $LD$ is then calculated for each set, and the average of these values represents the average $LD$ for that particular $c$.  The average $LD$ value thus computed, is plotted against the control parameter $c$, in Fig.~\ref{fig:fig1}b.

By comparing with the bifurcation diagram in Fig.~\ref{fig:fig1}a, we observe that $LD$ drops down suddenly at the first period-doubling bifurcation and is decreasing further at subsequent bifurcations till chaos is reached.  Then at every periodic window, it goes up again indicating a transition from chaos to periodicity.  For values of $c>50$, there is an increase in $LD$ till periodicity is reached.  From direct numerical simulations of equation(1) we find that beyond c=50, the dynamics is intermittently periodic cycles and for c> 80, the system settles to a periodic 1- cycle. We also show the value of $LD$ for purely periodic sine wave data, in the blue dotted line on top and that for pure stochastic Gaussian data in the black dotted line below for comparison.  The deviations in values that lie between these lines indicate the ranges of $LD$ between these limiting cases.

Additionally, we observe an interesting result that, not only do the mean values of $LD$ decrease in the chaotic regime but they also show an increase in the standard deviation(STD) of $LD$ for each value of control parameter $c$ within the same range.  This reflects the increase in the differences in $LD$ values derived from different segments of data for the same $c$ value.  We understand this as coming from the variations in the distribution of points on different parts of the chaotic trajectory and hence variations in their proximity patterns in the recurrences.  We find $LD$ is a good measure that can capture these variations and thus characterise the increased complexity in the geometry of the chaotic trajectory underlying the data.  We also report the STD in $LD$ can serve as a more effective measure to indicate transitions in dynamics between periodic and chaotic states.

We present the results of similar computations in Fig.~\ref{fig:fig1}b, zooming into a narrower range in $c$ within which the main transitions occur, highlighting the average $LD$ values at the transition points.  The critical values of $c$ at which transitions happen are indicated as insets, some of the values like $c=2.8286, 3.834, 4.1235,$ and $4.20408$ are reported earlier\cite{sarmah2013period}, while others shown are from our analysis of the trajectories and their bifurcations.

\begin{figure}[ht]
\centering
\includegraphics[width=0.5\textwidth]{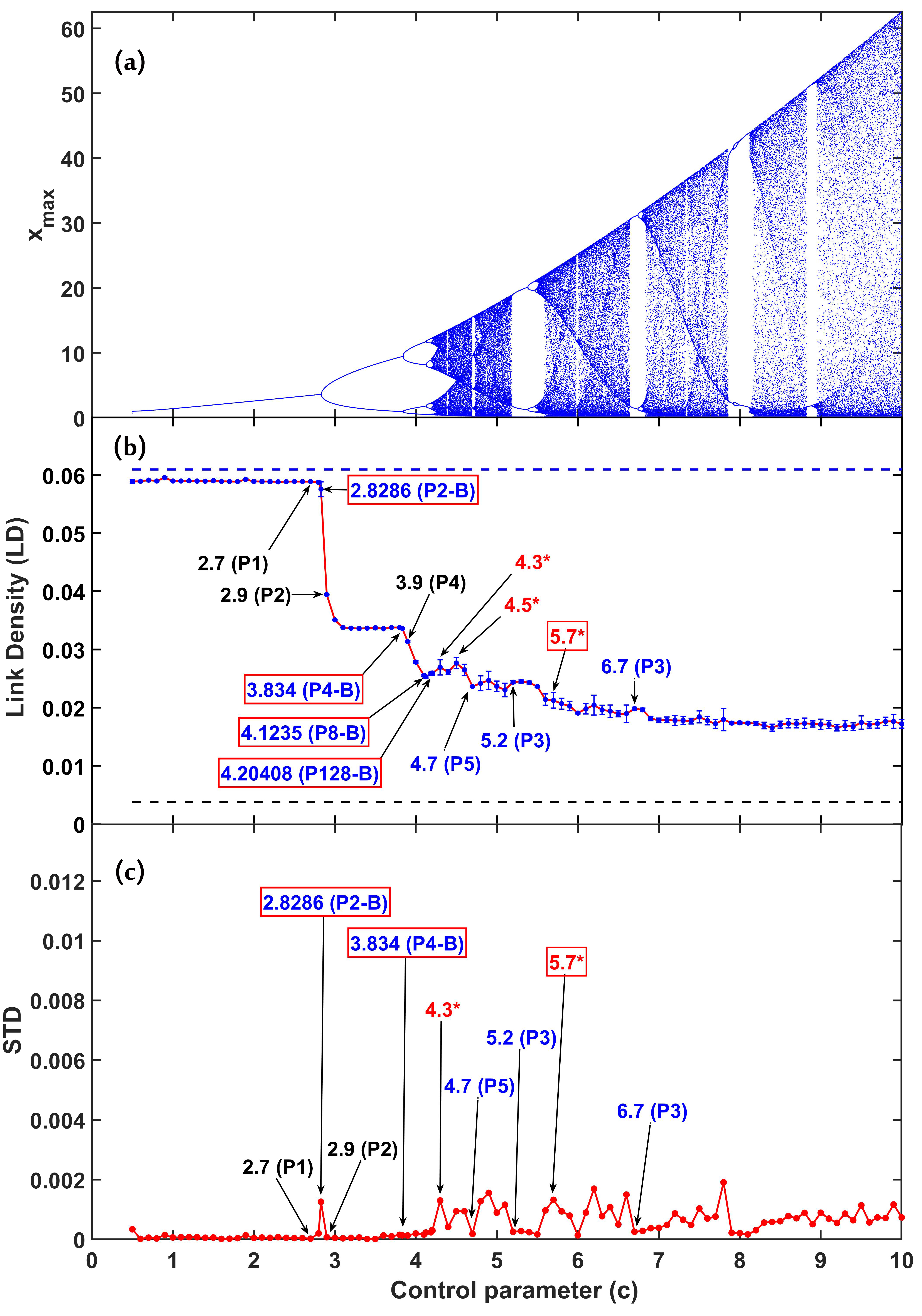}
\caption{ Transitions in the dynamics of the R\"ossler system for the parameter $c$ in the range from  $c$= 0.5 to 10 with $a=b=0.2$. (a) Bifurcation diagram indicating the period-doubling bifurcations to chaos and transitions to periodic windows within the chaotic region. (b) Average $LD$ for the same range of parameter $c$ and (c) Standard deviations in the $LD$ values corresponding to the average values in (b).  The critical values of $c$ at which transitions happen are indicated as insets, values associated with periodic dynamics are shown in black with cycle numbers in brackets and those for chaos in red with * above $c$ value.  The bifurcation points are shown in blue B after cycle numbers.}
\label{fig:fig2}
\end{figure}

From the analysis presented, it is clear that the average $LD$ exhibits higher values during the limit cycle phases of the  R\"ossler system and progressively decreases as bifurcations occur.  This trend indicates that during stable, periodic behaviors or limit cycles, the system's network structure is more interconnected, due to the regular and close recurrence of points in phase space.  Moreover, the standard deviation ($STD$) of $LD$ remains low, suggesting consistent network connectivity coming from a uniform distribution of points over the trajectory.  As the system transitions through bifurcations towards chaos, the average $LD$ diminishes, signifying a decrease in overall network connectivity that arise from irregular arrangements of points.  Then the $STD$ of $LD$ increases, indicating increased variability in network structure coming from nonuniform distribution of points over the trajectory.  These variations are seen clearly in Fig.~\ref{fig:fig2}b and Fig.~\ref{fig:fig2}c.

\section{\label{sec:level1}Dynamic transitions under slowly varying parameter}
In this section, we consider the impact of slowly varying parameter on the system's behavior, particularly in inducing transitions.  In real systems, it is possible that the parameters change due to external influences or internal variations.  Then the time scale at which the parameter varies and that of the system's inherent dynamics can be different.  This difference often plays a crucial role in the onset of transitions in the system \cite{tzou2015slowly}. We consider one such situation in the dynamics of the R\"ossler system, where the control parameter changes over a time-scale that is much longer than the system's dynamical time-scale\cite{tzou2015slowly,cantisan2023rate}.

For modeling the slowly varying parameter $c$, we choose its time-dependence as $c(t)$ given by
\begin{align}
c(t) = c_1 + \left( \frac{c_2 - c_1}{T} \right) \cdot (T_0 + t)
\label{eq:controlfunction}
\end{align}

Here, $c_1$ and $c_2$ represent parameter values included covering the different dynamical states of the system. In our computations, we fix $T = 300,000$, $T_0 = 1,000$, and increment $t$ by $0.05$. This would then mean that while the system dynamics is advanced in time steps of 0.05, the change in parameter in the same time step will be  $\approx 10^{-7}$ times slower.  Initially, we generate a dataset of 10,000 data points using $c_1$, discard the first 5,000 points to eliminate transients, and then generate 300,000 points as $c$ varies based on $c(t)$ as given in equation (\ref{eq:controlfunction}). We do a sliding window analysis of 3000 points per window with a slide of 10 on the remaining data points.  For each window after proper embedding, the recurrence network is generated and the $LD$ computed. The average values of $LD$ from 1000 consecutive windows and their standard deviations are then used in the analysis.  This method thus tracks the variation of $LD$ over time as the system and parameter evolve at differing time scales and assesses its sensitivity to transitions in the system. 

\begin{figure*}[ht]
\centering
\includegraphics[width=1\textwidth]{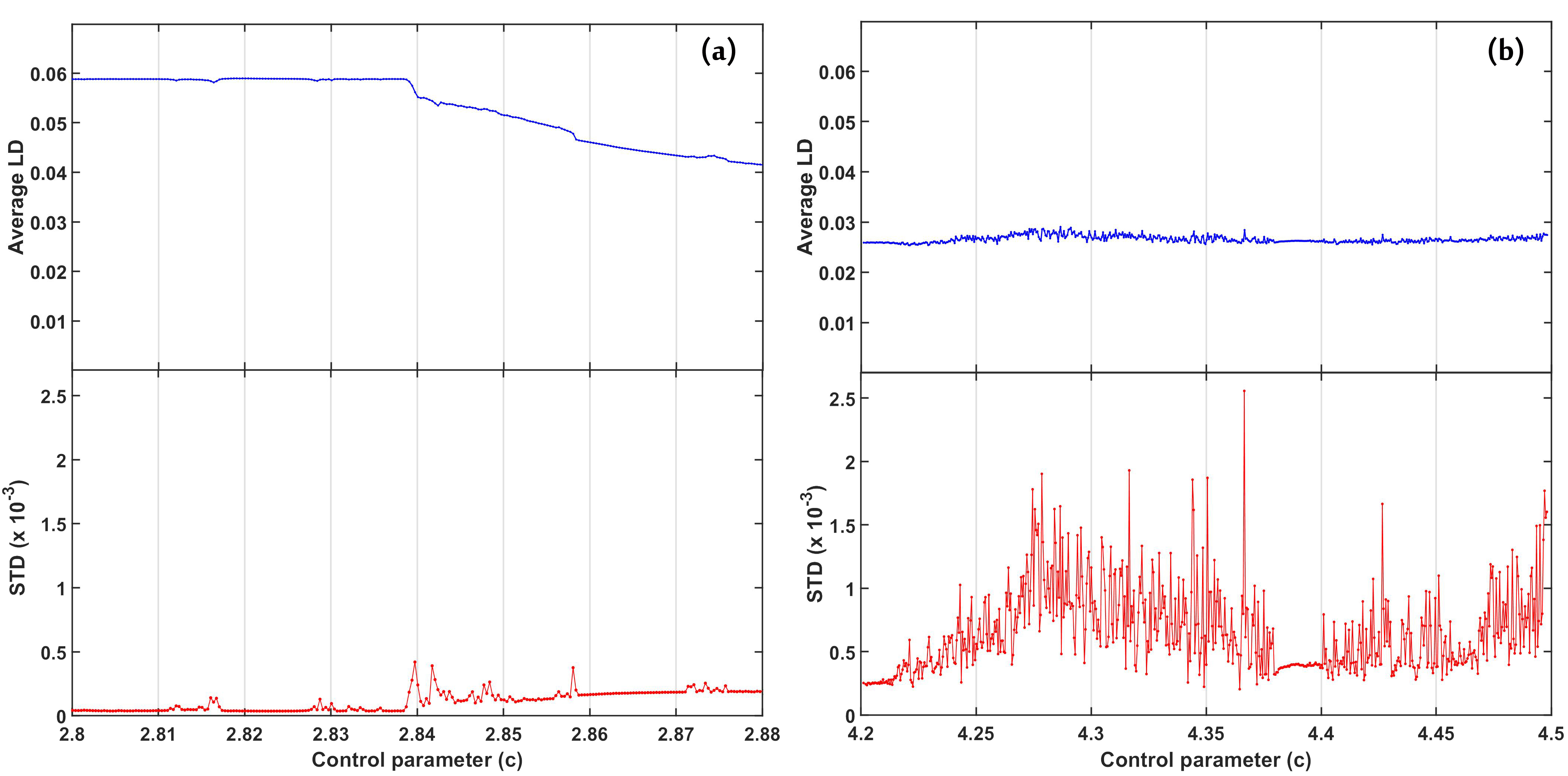}
\caption{Average $LD$ and its standard deviation computed using sliding window analysis on data generated from R\"ossler system with slowly varying parameter $c$ in the range (a)  from $c = 2.8$ to $2.88$  and (b)from $c = 4.2$ to $4.5$ .  The average is over 1000 consecutive windows and the values of $c$ shown correspond to their midpoint. }
\label{fig:fig3}
\end{figure*}

We present the results of this analysis for two ranges of $c$ values, the transition from a limit cycle to period two cycle through period doubling for  $c_1 = 2.8$ to $c_2 = 2.88$ in Fig.\ref{fig:fig3}a, and transition point from chaos to periodic and then back to chaos from $c_1 = 4.2$ to $c_2 = 4.5$ in Fig.\ref{fig:fig3}b. The parameter values shown correspond to the midpoint of 1000 windows used in averaging.  It is clear from Fig.\ref{fig:fig3}a, that there is a dip in average $LD$ and an increase in $STD$ of $LD$, as $c$ value reaches 2.838.  From our earlier analysis, we can say that this would correspond to the period-doubling transition point.  

The values of average $LD$ are less in Fig. \ref{fig:fig3}b compared to the values in Fig. \ref{fig:fig3}a, which indicates a state of chaos in Fig. \ref{fig:fig3}b.  Also, the variations in the $STD$ of $LD$ show more pronounced variations that indicate the region of chaos and the presence of periodic windows inside the chaotic region.  Thus, we see clear indications of transitions to chaos and transitions to periodic windows inside the chaotic region that happens in the system's dynamics due to slowly varying parameter $c$.  
\section{\label{sec:level1}Summary and Results}
In this study, we present a simple and efficient method to track dynamical transitions that happen in systems due to changes in their parameter values. We consider the changes in parameter made externally as well as the parameter changing slowly as the system evolves.  Taking the specific case of R\"ossler system, we illustrate how the recurrence network measure, $LD$, computed from time series data, can effectively detect the transition points in the system in both these cases.  By generating the time series corresponding to the bifurcation scenario in R\"ossler system, and computing the $LD$ from its recurrence network, we observe that $LD$ decreases suddenly at the first period-doubling bifurcation, followed by a further decrease at subsequent bifurcations till chaos is reached.  Near every periodic window, we find increases again indicating a transition from chaos to periodicity. This underscores the potential of $LD$ as a primary indicator for understanding and identifying the system's underlying dynamical transitions. 

We rely on recurrence network measures in this context since they capture the pattern of neighbourhood relationships in the phase space structure of a dynamical system.  Especially, we find the measure $LD$ shows good performance in detecting dynamical transitions since it quantifies the geometric pattern of trajectory points. It is computationally easy to get this measure from reconstructed phase space from data or time series of the system and is robust even with small data sets.  Moreover, when the dynamics is chaotic, due to the nonuniform distribution of points on the trajectory, the proximity of points changes locally. Then the values of $LD$ computed from data show increased standard deviation about the average value. Therefore, we propose the STD of $LD$ as an effective measure that can detect the points of transitions between periodic and chaotic dynamical states.

We also compute these measures, $LD$ and STD of $LD$ for the particular situation when the parameter is varying due to internal or intrinsic perturbations but very slowly compared to the system dynamics. In this case also, we find the variations in $LD$ and STD of $LD$ can detect transition points in the system.  The technique can be applied to a real-world complex system where a relevant quantity is being continuously monitored. The Link Density can then be computed in real-time and be used to track the dynamical transitions that the system is undergoing, which in turn may provide an early-warning signal. Moreover, the method adopted in this study can be easily integrated with machine learning for efficient analysis of transitions from data sets of complex systems.

\section{Acknowledgment}
RJ and KPH acknowledge the computing facilities at Inter-University Centre for Astronomy and Astrophysics (IUCAA), Pune. RJ acknowledges the HPC facility at \mbox{Rajagiri} School of Engineering and Technology (RSET).

\end{document}